\documentclass[aps,physrev,reprint,twocolumn,10pt,tightenlines,superscriptaddress,notitlepage,nofootinbib,longbibliography,floatfix]{revtex4-2}

\usepackage[T1]{fontenc}
\usepackage{lmodern}
\usepackage{amsmath,amssymb,mathtools}
\usepackage{graphicx}

\graphicspath{{../figures/}{figures/}}
\usepackage{xcolor}
\usepackage{microtype}

\usepackage[hidelinks]{hyperref}
\IfFileExists{arxiv_version/revtex_arxiv.bst}{%
  \newcommand{\arxivstylepath}{arxiv_version/}%
  \newcommand{\arxivbibfile}{references}%
}{%
  \newcommand{\arxivstylepath}{}%
  \newcommand{\arxivbibfile}{references}%
}

\begin{document}

\title{FNO-Based Inverse Co-Design of Semiconducting Quantum-Dot Devices}
\author{Marek Rychetsk\'y}
\affiliation{Department of Engineering, University of Oxford, Parks Road, Oxford OX1 3PJ, UK}
\author{Valentin Duruisseaux}
\affiliation{California Institute of Technology, Pasadena, CA 91125, USA}
\author{Freya Shah}
\affiliation{Department of Engineering, University of Oxford, Parks Road, Oxford OX1 3PJ, UK}
\author{David L. Craig}
\affiliation{QuantrolOx Ltd, 8 King Edward Street, Oxford, OX1 4HL, United Kingdom}
\author{Ming Ni}
\affiliation{Institute of Science and Technology Austria, Am Campus 1, 3400 Klosterneuburg, Austria}
\author{Georgios Katsaros}
\affiliation{Institute of Science and Technology Austria, Am Campus 1, 3400 Klosterneuburg, Austria}
\author{Taylor L. Patti}
\affiliation{NVIDIA Corporation, Santa Clara, CA, USA}
\author{Anima Anandkumar}
\affiliation{California Institute of Technology, Pasadena, CA 91125, USA}
\author{Natalia Ares}
\affiliation{Department of Engineering, University of Oxford, Parks Road, Oxford OX1 3PJ, UK}

\begin{abstract}
Gate-defined semiconductor quantum dots are promising building blocks for scalable spin-based quantum processors, however their operational properties depend on several fabrication and control parameters, such as lithographic gate layout and applied voltages. Existing device architectures have been refined through repeated fabrication and experimental testing, producing layouts that support reliable device operation. However, this process does not exhaustively explore the design space, limiting the ability to obtain qubits with targeted properties. We address these challenges by presenting a differentiable inverse co-design framework for electrostatically defined quantum dots that jointly optimises gate voltages and the gate electrode layout. A semiclassical Thomas--Fermi electrostatic simulator provides reference data for a Fourier neural operator (FNO) surrogate, which predicts the two-dimensional charge-density field. The desired spatial distribution of the confined charge defines the target quantum-dot geometry for inverse design. The FNO reproduces charge densities with a relative $L^2$ test error of just 0.6\% and enables millisecond-scale differentiable evaluations. Relative to the CPU-based physics simulator, the FNO achieves a 17,000-fold speed-up on GPU for a single design, with further performance gains enabled by batched inference. When embedded in a gradient-based design loop with fabrication constraints, the FNO can produce both vertically and horizontally elongated double-dot confinement geometries from the same initial conditions. The ability to engineer these distinct confinement configurations demonstrates control over quantum-dot shape and position, providing a route towards designing devices with distinct qubit-relevant operating properties prior to fabrication.
\end{abstract}

\maketitle

\section*{Introduction}

Semiconductor spin qubits encode quantum information in the spin states of electrons or holes, which can be confined in electrostatically defined quantum dots (QDs) \cite{Loss1998, Hanson2007, Chatterjee2021, Stano2022, Burkard2023}. Gate electrodes shape the electrostatic potential and thereby control the position, size, shape, and occupation of the QDs, as well as their coupling to charge reservoirs and neighbouring QDs. The QDs' nanometre-scale confinement, electrical tunability, and compatibility with semiconductor processing make QDs attractive building blocks for scalable quantum processors \cite{Zwerver2022,Neyens2024, Zhang2025, FernandezDeFuentes2026,HRL2026, Undseth2026Parity}. 
Holes confined in electrostatically defined QDs within Ge/SiGe heterostructures are a particularly promising platform for spin qubits, primarily due to their strong spin--orbit coupling, which enables fast electrical control and, together with their low effective mass, relaxes fabrication constraints, while maintaining low dephasing rates~\cite{Scappucci2021,Lodari2019,Hendrickx2020,Jirovec2021, Hendrickx2024, Stehouwer2025, Jirovec2026, Kelly2026, Dijkema2026}.

The complexity of operating semiconductor QD devices, which arises from multiple coupled gate electrodes, device variability, and disorder, has motivated the development of automated characterisation and control methods~\cite{Baart2016, Botzem2018, Mills2019, Zwolak2023Automation, Alexeev2025AI}. These include a range of machine-learning-based approaches, with recent advances demonstrating fully autonomous tuning from a de-energised device to coherent Rabi oscillations~\cite{Schuff2026}. Physics-aware machine learning has also enabled the inference of disorder potentials from experimental measurements~\cite{Craig2024RealityGap}, improving the predictive agreement between device simulations and experiments.

Post-fabrication tuning, however, can only explore the operating space supported by the lithographic layout. If the gate electrode geometry does not generate a suitable confinement landscape, favourable qubit properties, such as enhanced Rabi frequencies~\cite{Bosco2021, AbadilloUriel2023}, may remain inaccessible even when the available gate voltages are tuned optimally. This motivates a pre-fabrication framework that co-optimises gate geometry and voltages, expanding the accessible device design space beyond that achievable by tuning a fixed layout alone.

Despite these advances, pre-fabrication optimisation of QD gate electrode layouts remains largely unexplored \cite{Lee2025,Sanders2025}. Data-driven methods have been used to optimise lithographic fabrication processes \cite{Mei2021} and metasurface design~\cite{Jiang2019, Kudyshev2020}, and physics-informed machine learning has been applied to semiconductor heterostructure growth \cite{Shen2026}. For quantum point contacts (QPCs), algorithmic approaches have optimised electrostatic potentials for fixed gate geometries \cite{Chatzikyriakou2022,Benestad2025}. These advances highlight the potential of data-driven approaches for optimising device designs before fabrication.

Experimentally exploring the gate layout design space would require fabricating and measuring a sufficiently large number of devices, demanding substantial time and resources. For computational design, physics-based simulations provide a way to evaluate device properties for proposed geometries and operating conditions, but such models can be computationally expensive and may not provide gradients with respect to the design variables. This makes repeated evaluation and gradient-based optimisation difficult. Machine-learning surrogate models provide a practical route to address these limitations by learning the mapping between device design parameters and simulated responses, after which the learned model can be evaluated rapidly. When the surrogate is differentiable, gradients of the desired response can also be propagated back to the design variables, enabling direct optimisation through backpropagation \cite{Carleo2019,Molesky2018, Craig2024MasterEquation}.

Neural operators are a special class of machine-learning architectures designed to learn mappings between functions, which makes them suitable for physical problems in which both inputs and outputs are spatially varying fields \cite{Kovachki2023,Azizzadenesheli2024,Berner2026}. The most prominent example is the Fourier neural operator (FNO)~\cite{Li2021FNO,Duruisseaux2025FNO}, which represents these mappings using learned transformations of spatial Fourier modes, providing a global receptive field and enabling evaluation across different spatial discretisations. Here, we use the FNO as the differentiable forward model in our inverse-design pipeline for a double quantum dot (DQD) device.
We demonstrate the framework for holes in planar Ge, with the same approach extending naturally to other device architectures and material systems. The FNO's field-to-field formulation allows the model to capture the electrostatic potential's response while providing the differentiability and computational efficiency required for gradient-based design. This type of pipeline has been successfully applied across a broad range of applications, including the design of electromagnetic and photonic devices \cite{Augenstein2023,Gu2022NeurOLight,Zhu2024PACE,Duruisseaux2025Photonics}, and the control of quantum systems \cite{Pipi2026}.

\begin{figure*}[tp]

\centering

\includegraphics[width=0.96\linewidth]{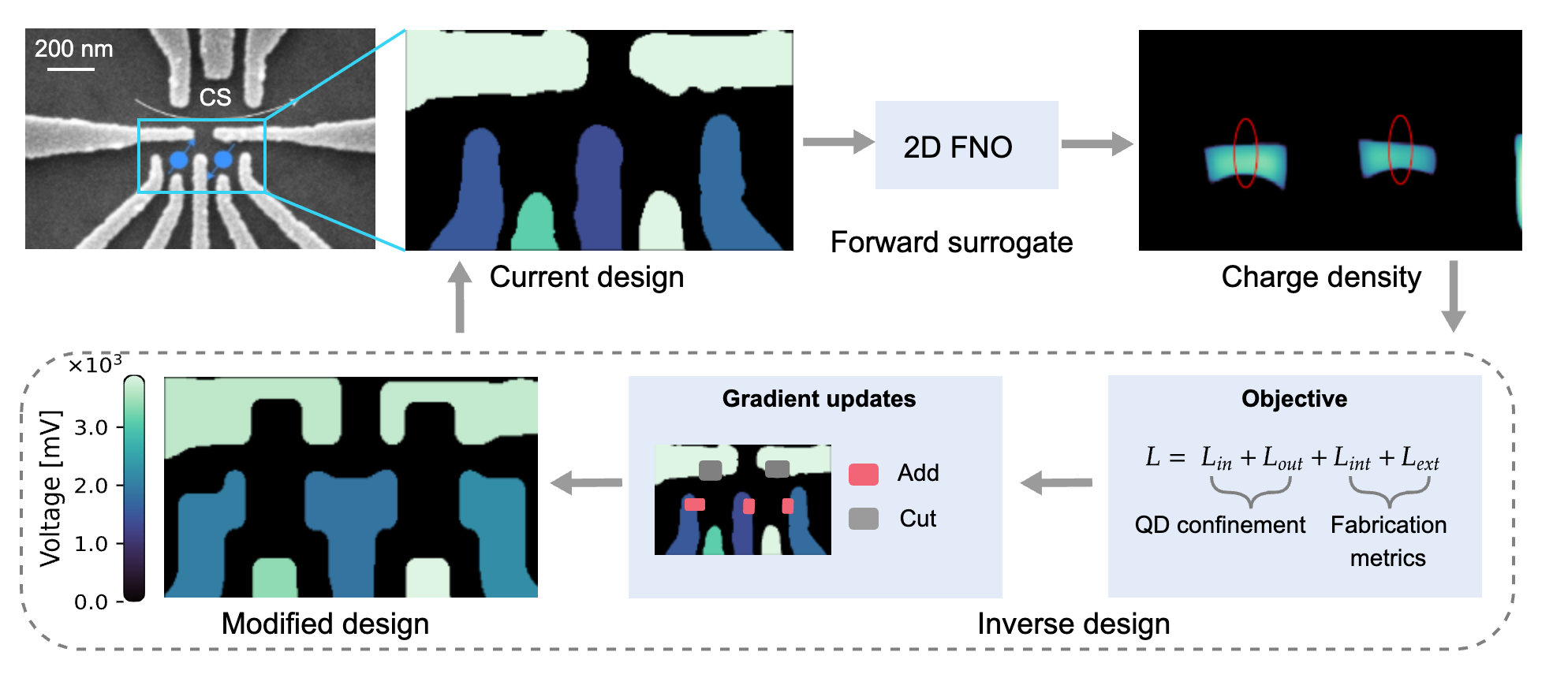}

\caption{\textbf{FNO-based quantum-dot inverse-design workflow.} A semiclassical electrostatic Thomas–Fermi model generates reference two-dimensional charge-density maps from sampled gate geometries, voltages, and fixed device parameters. The generated data is used to train a Fourier neural operator (FNO), providing a fast and differentiable surrogate for the reference simulator. The FNO is then embedded within the inverse-design optimisation loop, where the initial gate geometry and voltages are extracted from established designs (e.g., geometry extracted from an SEM image of a fabricated device that is combined with the correspondingly used gate voltages) as the starting configuration. Within this loop, the FNO  predicts the resulting charge-density distribution, enabling gradient-based updates of both the parameterised gate-edge geometries and gate voltages to match a prescribed target density (red ovals) while enforcing fabrication constraints.}

\label{fig:1}

\end{figure*}

Our workflow is shown in Fig. \ref{fig:1}. We generate reference charge-density maps using a semiclassical electrostatic Thomas--Fermi model, and use the distribution of the resulting confined charge to define the target QD geometry for the design objective. We train an FNO surrogate to approximate and accelerate the mapping from gate-electrode geometries and applied voltages to the corresponding charge-density fields, and embed the trained model within a gradient-based optimisation loop. The design variables parameterise local modifications to the gate-electrodes together with the applied gate voltages, with all continuous design variables optimised subject to explicit fabrication constraints.

This work's main objective is to realise controlled bespoke QD confinement geometries. Different confinement geometries can, in turn, influence qubit-relevant properties. In the present work, we therefore use the confinement profile as a directly accessible design target through which these properties may be influenced, without modelling them explicitly. Future extensions incorporating the required quantum-mechanical models could instead optimise properties such as g-factor and inter-dot coupling directly.

\section*{FNO surrogate for quantum dot devices}

Gradient-based inverse design requires a differentiable forward model that predicts the charge-density distribution from the gate electrode layout and applied voltages, while providing gradients of the resulting response with respect to these design variables. Computational efficiency is equally important, as the forward model must be evaluated repeatedly throughout the optimisation process.

For the inverse-design problem considered here, the device can be accurately described using a semiclassical numerical model~\cite{Craig2024RealityGap, Lidiak2025}. We can use this model to simulate the electrostatic potential and resulting hole density of a Ge/SiGe double QD device. The local hole density can be calculated using a two-dimensional Thomas--Fermi approximation, with the spatial distribution of the confined hole density defining the QD confinement shape used to guide the optimisation.

We construct the electrostatic model from the Ge/SiGe double QD device described in Ref.~\cite{Jirovec2021}, using the reported heterostructure and gate layout characteristics. The same modelling approach was used in Ref. ~\cite{Jirovec2022} to connect the electrostatic confinement landscape to the experimentally measured $g$-factor response. The device parameters, electrostatic boundary conditions and calibration procedure used here are detailed in \hyperref[sec:methods]{Methods}. Although this modelling approach provides the required mapping from gate-electrode geometry and applied voltages to the resulting hole density, it does not directly provide gradients of the converged density with respect to the gate geometry and voltages, as required for gradient-based inverse design. Moreover, the computational cost of conventional numerical methods makes their repeated evaluation increasingly prohibitive within an inverse-design loop. A typical simulation on the $343\times399$ pixel grid used in this work, which corresponds to $510\times593$ $\mathrm{nm}$, requires approximately 24~s, with the computational cost increasing substantially for larger spatial domains.

To overcome these limitations, we train an FNO on data generated by this physics-based simulator and subsequently use the trained model as a fast and differentiable surrogate for repeated evaluations of the expensive, non-differentiable simulator. It acts directly on the spatial fields and uses learned global spectral convolutions to communicate information across the simulation domain \cite{Li2021FNO,Kovachki2023}. This global receptive field is well matched to gate-defined electrostatics, where a local change in geometry or voltage can alter the confinement over a much larger region. After training, the neural operator is used as a fixed differentiable forward model, allowing gradients of the predicted density with respect to the input field to be computed by backpropagation and used directly in the inverse-design optimisation.

The FNO takes as input a two-dimensional field $a(x,y)$ defined on the device simulation grid, where $x$ and $y$ are the in-plane spatial coordinates. Each pixel of $a(x,y)$ encodes the normalised voltage applied to the gate electrode at that location, with a fixed background value assigned to regions outside the gates. The FNO maps this input field to the corresponding two-dimensional charge-density field,
\begin{equation}
\rho_{\mathrm{FNO}}(x,y)=\mathcal{F}_{\phi}[a](x,y).
\end{equation}

We generate the training data using the physics-based solver by sampling variations in gate-electrode geometry and applied voltages with respect to the double-QD reported in Ref.~\cite{Jirovec2021}, while keeping the other simulator parameters fixed.

The dataset intentionally spans a wider geometry and voltage range than is experimentally accessible, so that the surrogate remains reliable for intermediate configurations encountered during optimisation. Our dataset contains 30,000 training configurations and an independent 6,000-configuration test set. The detailed geometry generation, voltage sampling, FNO input encoding, model architecture and training procedure are described in \hyperref[sec:methods]{Methods}, with the numerical model and training parameters provided in Supplementary Table~S1.

\begin{figure*}[tp]

\centering

\includegraphics[width=\linewidth,height=0.72\textheight,keepaspectratio]{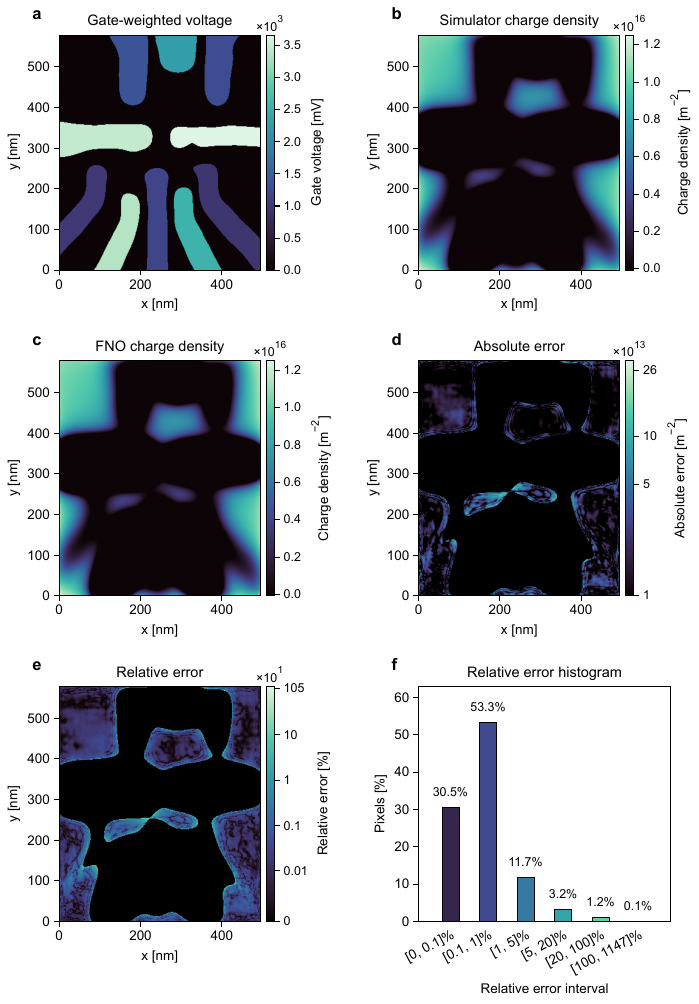}

\caption{\textbf{FNO prediction accuracy for a representative test configuration.} \textbf{a} Gate-weighted voltage field encoding the electrode geometry and applied voltages used as the FNO input. \textbf{b} Reference charge-density distribution calculated using the physics-based simulator. \textbf{c} Corresponding FNO-predicted charge-density distribution. \textbf{d} Pixelwise absolute error between the reference and FNO predictions. \textbf{e} Pixelwise relative error. Pixels with an absolute error below $10^{13}\,\mathrm{m}^{-2}$ are masked in both \textbf{d} and \textbf{e}. \textbf{f} Distribution of the relative error after applying the absolute-error threshold. The largest discrepancies are localised near sharp charge-density boundaries, with the overall spatial structure of the reference charge density closely reproduced by the prediction.}

\label{fig:2}

\end{figure*}

The FNO achieves a relative $L^2$ error of just 0.6\% across the independent test set, with a mean absolute error of approximately $2.5\times10^{13}\,\mathrm{m}^{-2}$. Figure \ref{fig:2} illustrates the agreement for a representative test configuration. The gate geometry and applied voltages are combined into the gate-weighted input field shown in Fig. \ref{fig:2}a. For this input, the physics-based simulator produces the reference charge-density distribution in Fig. \ref{fig:2}b, while the FNO predicts the density shown in Fig. \ref{fig:2}c. The two density distributions show close agreement across the full spatial domain, reproducing the position and shape of the confined regions as well as the narrow density features surrounding the double QD. To resolve differences that are not apparent from the density maps, we compare the pixelwise absolute and relative errors in Fig. \ref{fig:2}d and \ref{fig:2}e. In both maps, pixels with an absolute error below $10^{13}\,\mathrm{m}^{-2}$ are masked because this error is negligible compared to the characteristic hole density values of order $10^{16}\,\mathrm{m}^{-2}$. The remaining absolute error is concentrated predominantly along sharp density boundaries, where small differences in the predicted boundary position produce the largest pixelwise deviations. In this example, the absolute error reaches approximately $3\times10^{14}\,\mathrm{m}^{-2}$ at these boundaries, remaining well below the characteristic hole density values.

The relative-error distribution in Fig. \ref{fig:2}f further shows that 83.8\% of the unmasked pixels have errors below 1\%, while 95.5\% have errors below 5\%. Larger relative errors occur only for a small fraction of pixels and are concentrated near density boundaries. In these regions, a small absolute difference in the predicted and reference boundary can correspond to a large relative error because the local reference density is small. The close agreement of the density fields and the low test-set error demonstrate that the FNO accurately reproduces the physical-simulator response required for the subsequent inverse-design optimisation.

The trained FNO reduces the cost of a single density evaluation by four orders of magnitude. All timings were measured on a single node equipped with two Intel Xeon Platinum 8480C CPUs ($2\times56$ physical cores) and NVIDIA H100 80\,GB SXM GPUs. The self-consistent Thomas--Fermi solver, implemented as a serial CPU code, requires $24.4\,\mathrm{s}$ per design on a single core. In contrast, on one H100, the FNO predicts the corresponding hole density in $1.42\,\mathrm{ms}$, yielding a $\sim17{,}000\times$ speed-up. Batching $26$ designs per forward pass further reduces the amortised cost to $1.03\,\mathrm{ms}$ per design, corresponding to a $\sim24{,}000\times$ speed-up. The surrogate thus reduces device-evaluation times from tens of seconds to the millisecond scale: millions of designs can be evaluated within an hour using the FNO, when it would take more than a year using the serial solver. This comparison should however be nuanced, since the baseline solver uses neither multicore parallelism nor GPU acceleration, and part of the observed gain therefore stems from hardware acceleration rather than the surrogate model alone. Nevertheless, the FNO remains substantially faster under CPU-only comparisons, achieving an $82\times$ speed-up on a single core and a $309\times$ speed-up when using all $112$ CPU cores.

\section*{Differentiable inverse design of quantum dot devices}

\subsection*{Formulation}

We apply the inverse-design framework to produce two deliberately different QD confinement profiles: vertically and horizontally elongated ellipses. These targets are chosen to represent clearly different confinement regimes and thereby produce distinct qubit-relevant properties, such as g-tensors. Starting from the double-QD device reported in Ref.~\cite{Jirovec2021}, we optimise the gate-electrode geometry and applied voltages to realise prescribed QD confinement geometries.

\begin{figure*}[tp]

\centering

\includegraphics[width=\linewidth,height=0.72\textheight,keepaspectratio]{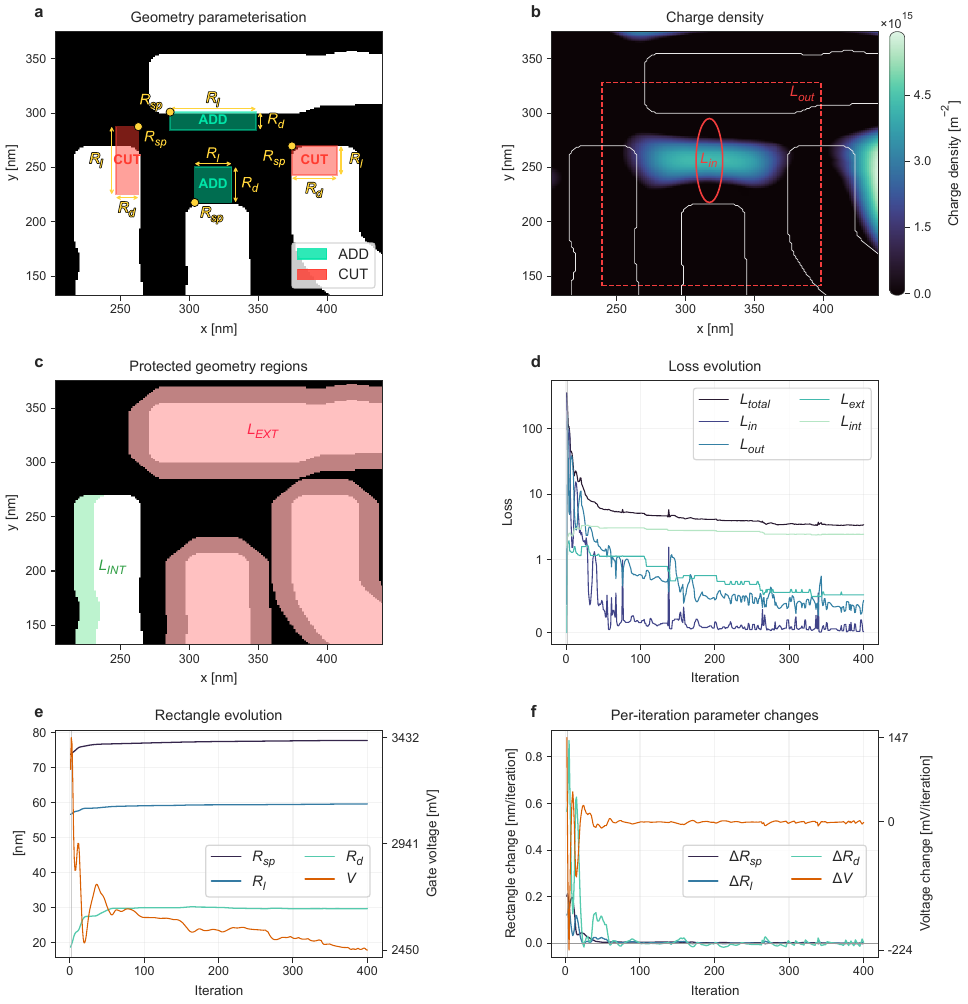}

\caption{\textbf{Gate-electrode parameterisation, optimisation objectives and convergence.}
\textbf{a} Rectangular edge transformation (RET) defined by the starting position $R_{\mathrm{sp}}$, tangential length $R_l$ and oriented depth $R_d$, allowing local expansion, contraction or no modification of the gate-electrode geometry.
\textbf{b} Target elliptical confinement region and surrounding rectangular region outside the target ellipse used to define the charge-density losses $L_{\mathrm{in}}$ and $L_{\mathrm{out}}$, respectively.
\textbf{c} Internal and external protected regions used to enforce fabrication constraints. The internal loss $L_{\mathrm{int}}$ penalises insufficient gate-electrode coverage to maintain a minimum electrode width, while the external loss $L_{\mathrm{ext}}$ penalises violations of the minimum spacing between neighbouring electrodes.
\textbf{d} Evolution of the total loss and its charge-density and fabrication-constraint components during optimisation.
\textbf{e} Evolution of the gate-electrode geometry and voltage parameters during optimisation.
\textbf{f} Changes in the corresponding parameters between successive optimisation iterations.}

\label{fig:3}

\end{figure*}

For each QD, we select the gate electrode edges that define the confinement region and parameterise each edge modification using three geometric variables.
$R_{\mathrm{sp}}$, $R_l$ and $R_d$ specify the starting position, tangential length and oriented depth of the rectangular modification, respectively, as illustrated in Fig. \ref{fig:3}a. The comparison between the current density state and desired density state determines whether the selected gate-electrode geometry is locally expanded, contracted or left unchanged. Each participating gate electrode also has a trainable voltage $V$. 
We collect these trainable parameters in the design vector $\boldsymbol{\theta}=(R_{\mathrm{sp}},R_l,R_d,V)$. A gate electrode shared by multiple QDs retains a single voltage variable, while separate local geometry modifications act on its different QD-facing edges.

For each set of design variables, the parameterised gate electrode layout and voltages are passed to the trained FNO, which predicts the corresponding charge-density distribution. The total loss quantifies both deviation from the target confinement and violations of the imposed fabrication constraints. Because the geometry construction and FNO are differentiable, gradients of this loss are backpropagated to the continuous geometry and voltage parameters, which are optimised simultaneously.

\subsection*{Loss function}

To encourage agreement between the predicted charge-density profile and the target QD confinement while maintaining fabrication-compatible gate electrode geometries, we define the inverse-design objective as 
\begin{equation}
\label{eq:eq2}
\mathcal{L}
=
w_{\mathrm{in}}\mathcal{L}_{\mathrm{in}}
+
w_{\mathrm{out}}\mathcal{L}_{\mathrm{out}}
+
w_{\mathrm{ext}}\mathcal{L}_{\mathrm{ext}}
+
w_{\mathrm{int}}\mathcal{L}_{\mathrm{int}}.
\end{equation}

The density terms $\mathcal{L}_{\mathrm{in}}$ and $\mathcal{L}_{\mathrm{out}}$ govern the target confinement profile as shown in Fig. \ref{fig:3}b, while the fabrication terms constrain the evolving gate electrode layout as illustrated in Fig. \ref{fig:3}c. 
As shown in Fig. \ref{fig:3}b, the inner loss $\mathcal{L}_{\mathrm{in}}$ promotes a confined region within the target ellipse, while the outer loss $\mathcal{L}_{\mathrm{out}}$ suppresses the charge density values within the surrounding rectangle outside the target ellipse. These terms drive the charge density boundary towards the prescribed ellipse and thereby control both the orientation and size of the QD.
The external loss $\mathcal{L}_{\mathrm{ext}}$ enforces a minimum separation between neighbouring gate electrodes by penalising overlap of a selected electrode with the protected region surrounding an adjacent electrode, whereas the internal loss $\mathcal{L}_{\mathrm{int}}$ enforces a minimum gate-electrode width by penalising geometry changes that reduce the electrode width beyond the protected region. Their relative contributions are controlled by the weights $w_{\mathrm{in}}$, $w_{\mathrm{out}}$, $w_{\mathrm{ext}}$, and $w_{\mathrm{int}}$. Together, these terms direct the optimiser towards geometry and voltage changes that improve the target confinement while satisfying the imposed fabrication constraints.

Because the charge density varies continuously, the boundary of a QD is not uniquely defined. We therefore use a density threshold of $10^{14}\,\mathrm{m}^{-2}$ to define a confined region. During backpropagation, a straight-through estimator supplies a smooth gradient through the otherwise non-differentiable charge density threshold. The full implementation is described in Methods.

\section*{Inverse-designed double quantum dots}

\begin{figure*}[tp]
  \centering
  \includegraphics[width=\linewidth,height=0.72\textheight,keepaspectratio]{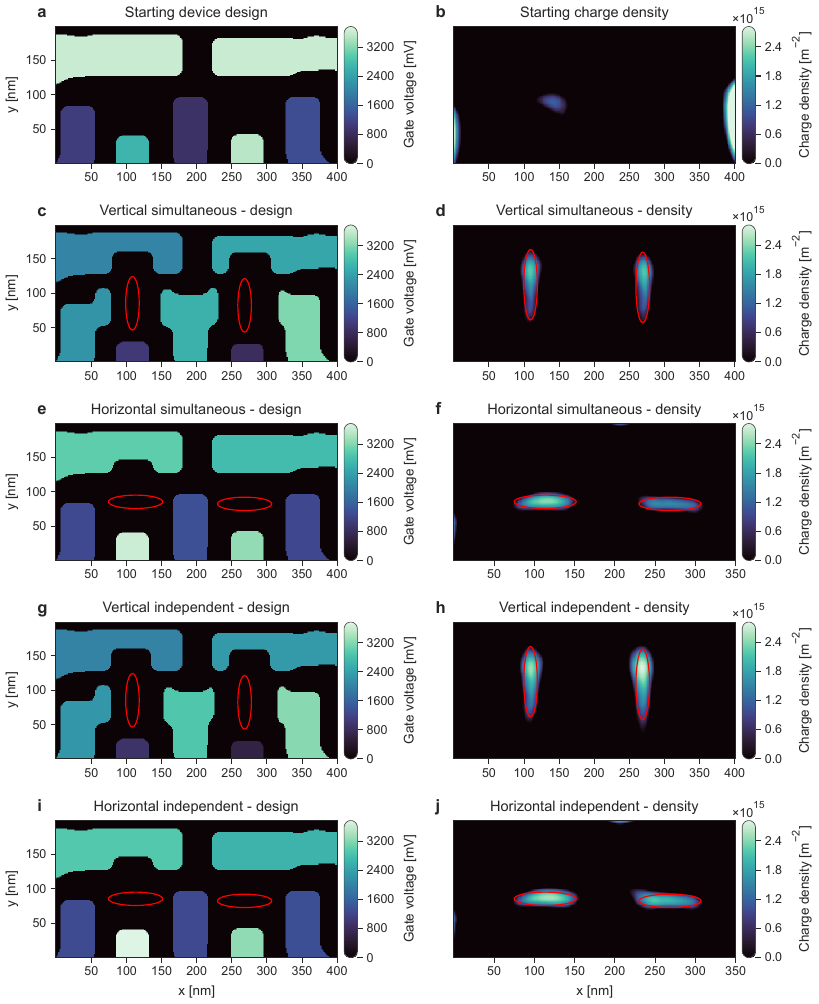}
  \caption{\textbf{Inverse design of double quantum dots with vertical and horizontal confinement orientations.} \textbf{a,b}, Initial preprocessed gate-electrode geometry, applied voltages and resulting charge-density distribution. \textbf{c,d}, Optimised gate-electrode geometry and charge-density distribution for vertically oriented confinement targets obtained by simultaneous optimisation of both QDs. \textbf{e,f}, Optimised gate-electrode geometry and charge-density distribution for horizontally oriented targets obtained by simultaneous optimisation. \textbf{g,h}, Optimised gate-electrode geometry and charge-density distribution for vertically oriented targets obtained by independent optimisation of each QD. \textbf{i,j}, Optimised gate-electrode geometry and charge-density distribution for horizontally oriented targets obtained by independent optimisation. Red ellipses indicate the prescribed target confinement regions.}
  \label{fig:4}
\end{figure*}

We apply the inverse-design framework to produce two deliberately contrasting confinement profiles, consisting of vertically and horizontally elongated QDs. The different orientations are chosen to realise distinct confinement geometries that can influence qubit-relevant properties. Figure~\ref{fig:4} compares the initial preprocessed gate-electrode geometry and applied voltages (Fig.~\ref{fig:4}a), together with the corresponding charge-density distribution (Fig.~\ref{fig:4}b), with the configurations obtained through inverse design. Starting from the same initial device configuration, the framework jointly optimises local gate-edge modifications and applied voltages to reshape the charge-density distribution towards the prescribed confinement profiles while satisfying the imposed fabrication constraints. We first optimise both QDs simultaneously for each target orientation (Fig.~\ref{fig:4}c--f) and then investigate whether independent optimisation of the two QDs yields comparable gate-electrode geometries, operating voltages and confinement profiles under identical inverse-design hyperparameters (Fig.~\ref{fig:4}g--j).

In simultaneous optimisation, the geometry parameters and voltages associated with both QDs are optimised together, with the total objective given by
\begin{equation}
  L_{\mathrm{T}} = L_{\mathrm{LQD}} + L_{\mathrm{RQD}},
\end{equation}
where $L_{\mathrm{LQD}}$ and $L_{\mathrm{RQD}}$ denote the loss functions associated with the left and right QDs, respectively, each defined as in Eq.~\ref{eq:eq2}. Starting from the same initial preprocessed gate-electrode geometry and voltage configuration, in independent optimisation, the left and right QDs are each optimised separately. For each QD, only the design of its four surrounding gate-electrode edges and the corresponding gate voltages are optimised, while the remaining gate electrodes retain their initial geometries and voltages. The loss associated with the other QD is excluded, such that $L_{\mathrm{RQD}}=0$ when optimising the left QD and $L_{\mathrm{LQD}}=0$ when optimising the right QD. Neither optimisation uses information from the other. The independently optimised gate-electrode geometries and voltages are subsequently combined to form the complete double-QD device. Since the shared gate electrode is assigned different voltages by the two independent optimisations, their mean is used when combining the local designs. The same inverse-design hyperparameters are used for both optimisation approaches, as indicated in Supplementary Table~S2.

For simultaneous optimisation, the resulting charge-density boundaries closely follow the prescribed confinement profiles, with the charge density extending over $99.1\%$ and $99.6\%$ of the target elliptical regions for the left and right QDs in the vertical design, respectively, and $98.5\%$ and $99.4\%$ in the horizontal design. The charge density outside the target ellipses remains below $1.9\%$ of the surrounding area for all four QDs. The optimisation therefore reproduces the prescribed confinement orientations with only small deviations around the target boundaries. Independent optimisation produces similarly close agreement with the prescribed confinement profiles. The charge density extends over $98.9\%$ and $99.6\%$ of the target elliptical regions for the left and right QDs in the vertical design, respectively, and $98.8\%$ and $99.2\%$ in the horizontal design. These values differ from those obtained by simultaneous optimisation by at most $0.4$ percentage points, while the charge density outside the target ellipses remains below $1.7\%$ of the surrounding area for all four QDs. 

The gate-electrode geometries obtained by combining the results of the independent optimisations closely match those obtained by simultaneous optimisation, with $95.1\%$ agreement for the vertical design and $100\%$ for the horizontal design, measured by the overlap of modified pixels. The largest differences between the two optimisation approaches occur on the gate electrode shared by the two QDs, which accounts for approximately $67\%$ of the differing geometry pixels in the vertical design. During independent optimisation, each QD modifies a different QD-facing edge of the shared gate electrode. These modifications are combined directly. However, for other gate-electrode layouts, the modifications could be incompatible, requiring additional geometric constraints to ensure a valid layout. Unlike the local gate-edge modifications, the voltage applied to the shared gate electrode is a single parameter, which can take different values when the two QDs are optimised independently. 

Averaging the independently optimised shared-gate voltage modifies the resulting charge-density distributions. For the horizontal design, the charge density outside the target ellipse decreases by $0.2$ percentage points for the left QD and increases by $0.4$ percentage points for the right QD, while the fraction of the target elliptical region covered by the charge density decreases by $1.9$ percentage points and increases by $0.6$ percentage points, respectively. For the vertical design, the charge-density region outside the target ellipse decreases by $0.5$ percentage points for the left QD and increases by $0.6$ percentage points for the right QD. The corresponding fraction of the target elliptical region covered by the charge density decreases by $5.2$ percentage points for the left QD and remains unchanged for the right QD.

To verify the final inverse-designed configurations, we evaluate their optimised gate geometries and voltages using the original physics-based Thomas--Fermi simulator. The resulting density profiles are visually indistinguishable from the FNO predictions and retain the prescribed horizontal and vertical confinement orientations. The close agreement between the FNO and physics-based simulations demonstrates that the FNO accurately reproduces the designed confinement profiles, with typical boundary differences of only $1$--$2$ pixels, corresponding to approximately 1.5--3 nm. This discrepancy is comparable to or smaller than the boundary variations produced by realistic uncertainties in physical device parameters. Such uncertainties arise from growth-induced variations in heterostructure layer thicknesses, which affect the 2DHG depth, as well as from fabrication-induced deviations of the gate-electrode geometry from the designed lithographic layout.

\section*{Conclusions}

We have developed a differentiable inverse-design framework that jointly optimises local gate-electrode geometries and applied voltages for semiconductor QD devices. Using Ge/SiGe double QDs as a case study, we train a Fourier neural operator (FNO) on charge-density distributions generated by a physics-based Thomas--Fermi simulator. The trained FNO provides a fast, differentiable surrogate for inverse design, reproducing the reference charge-density fields with a relative $L^2$ error of approximately $0.6\%$ and reducing the evaluation time from approximately $24\,\mathrm{s}$ to $1\,\mathrm{ms}$. Starting from the same initial device configuration, the framework realises vertically and horizontally oriented elliptical confinement profiles, with the charge density extending over more than $98\%$ of each target ellipse and covering less than $2\%$ of the surrounding regions, while satisfying the imposed gate-electrode spacing and minimum-width constraints. The resulting confinement shapes are expected to remain robust against moderate fabrication and modelling uncertainties, with residual deviations amenable to further correction through post-fabrication gate-voltage tuning. These results demonstrate control over distinct QD confinement landscapes through joint optimisation of gate-electrode layouts and operating voltages.

More broadly, our approach shifts part of QD device optimisation from post-fabrication tuning to pre-fabrication design. Rather than relying on voltage tuning to explore the operating regimes supported by a fixed gate architecture, the gate-electrode geometry itself is optimised to produce a prescribed confinement landscape. This provides a route towards designing devices with targeted qubit-relevant characteristics before fabrication, reducing reliance on empirical design iterations. While the present framework uses the QD confinement geometry as the design objective, future extensions incorporating more advanced simulators could directly optimise properties such as $g$-factor anisotropy, spin--orbit coupling and inter-dot coupling.

Extending the framework to larger QD arrays presents additional challenges, as generating full-device simulation datasets becomes increasingly computationally expensive. The close agreement between independent and simultaneous optimisation of the two QDs suggests a route towards local or modular inverse design. Individual QDs could be simulated and optimised using local surrogate models, allowing multiple confinement regions to be optimised in parallel before combining gate-electrode modifications into a larger device layout. Such an approach could reduce the computational cost and complexity of generating training datasets for larger QD arrays. The field-based formulation of neural operators is well suited to developing these local surrogates. Further work is also needed to determine effective gate architectures for larger QD arrays, including how many gate electrodes are required and how to preserve local controllability. Together, these developments represent a step towards designing QD hardware for desired physical parameters before fabrication and towards complex QD arrays in which reproducible local device properties are essential for operation.

\section*{Methods}
\label{sec:methods}

\subsection*{Physics-based electrostatic and Thomas--Fermi model}

The semiclassical simulator maps a specified gate layout and applied gate voltages to the electrostatic potential and resulting charge density in the two-dimensional hole gas (2DHG). To calculate the charge density, we first determine the electrostatic potential in the 2DHG plane. The total potential contains contributions from the surface gate electrodes $\Phi_{\mathrm{gate}}$, a background surface potential $\Phi_{\mathrm{surf}}$, and the screening potential $\Phi_{\mathrm{scr}}$ arising from the self-consistent electrostatic response of the accumulated holes,

\begin{equation}
\Phi_{\mathrm{tot}}(x,y)=\lambda\Phi_{\mathrm{gate}}(x,y)+\Phi_{\mathrm{surf}}+\Phi_{\mathrm{scr}}(x,y).
\end{equation}
Here $\Phi_{\mathrm{gate}}$ and $\Phi_{\mathrm{scr}}$ are calculated from the gate geometry and carrier distribution, respectively, while $\Phi_{\mathrm{surf}}$ is a spatially uniform background surface potential calibrated to reproduce the experimentally reported background charge density. The empirical scale factor $\lambda$ accounts for the effective strength of the gate electrode induced potential and is calibrated against the configuration in Ref. ~\cite{Jirovec2021}, as described below. The calculation of the gate electrode induced potential, carrier screening and charge density follows the electrostatic and self-consistent treatment of Nixon and Davies \cite{NixonDavies1990}.

For the gate electrode induced contribution, each patterned electrode is decomposed into rectangular segments. For a rectangular segment held at voltage $V$, the potential is obtained from the contributions of its four corners using
\begin{equation}
g(u,v,d)=\frac{1}{2\pi}\tan^{-1}\!\left[\frac{uv}{d\sqrt{u^2+v^2+d^2}}\right],
\end{equation}
where $u$ and $v$ are the in-plane distances from the evaluation point to the corresponding electrode corner along the two coordinate axes, and $d$ is the vertical separation between the surface gate electrode and the 2DHG. The contributions from all rectangular segments and electrodes are summed to obtain $\Phi_{\mathrm{gate}}(x,y)$.

For a charge density $\rho(\mathbf{r})$, the screening potential is evaluated as a 2D Coulomb convolution,
\begin{equation}
\Phi_{\mathrm{scr}}(\mathbf{r})=\int G_C(\mathbf{r}-\mathbf{r}',d,\varepsilon_r)e\rho(\mathbf{r}')\,\mathrm{d}^2\mathbf{r}'.
\end{equation}
The local charge density is calculated semiclassically using the 2D Thomas--Fermi approximation,
\begin{equation}
\rho(\mathbf{r})=D_{2\mathrm{D}}\left[E_F-e\Phi_{\mathrm{tot}}(\mathbf{r})\right]_+,\qquad D_{2\mathrm{D}}=\frac{g m^*}{2\pi\hbar^2}.
\end{equation}
Here, $D_{2\mathrm{D}}$ is the two-dimensional charge density of states, $[\cdot]_+$ denotes the positive part of the enclosed expression, ensuring a non-negative charge density, $E_F$ is the effective Fermi level, $m^*$ is the in-plane hole effective mass, and $g$ is the effective degeneracy. Because $\rho$ determines $\Phi_{\mathrm{scr}}$ while $\Phi_{\mathrm{tot}}$ determines $\rho$, the charge density and potential are iterated self-consistently with under-relaxation until convergence.

\subsection*{Device parameters and calibration}

We parameterise the simulator using the measured planar Ge/SiGe double-QD device of Ref.~\cite{Jirovec2021}. The gate electrode-to-2DHG depth is $d=43\,\mathrm{nm}$ from the reported heterostructure layer stack, the Ge relative permittivity is $\varepsilon_r=16.2$ \cite{Hofmann2019GeSiGe}, and the in-plane hole effective mass is $m^*=0.054m_0$, consistent with calculations for the same device reported in Ref.~\cite{JirovecThesis2021}. We set the density-of-states degeneracy to $g=2$ for the twofold-degenerate lowest heavy-hole subband.

We calibrate the spatially uniform surface potential to $\Phi_{\mathrm{surf}}\approx-459\,\mathrm{mV}$ so that the simulator reproduces the reported background charge density of $9.7\times10^{11}\,\mathrm{cm}^{-2}$. We set the gate-voltage scale factor to $\lambda\approx0.706$ by comparing the simulated charge configuration with the experimentally reported gate voltages in Ref.~\cite{Jirovec2021}. We choose the largest $\lambda$ that retains two separated QDs together with a separated charge-sensor dot at these voltages. Decreasing $\lambda$ reduces the QD size and produces more symmetric lateral confinement, which is inconsistent with the strongly anisotropic confinement expected from the experimentally observed $g$-factor values.

To assess the sensitivity of the inverse-designed confinement profiles to uncertainty in the physical model, we re-evaluate the final designs while varying the 2DHG depth and in-plane effective mass around their baseline values. A change in the 2DHG depth of $1\,\mathrm{nm}$ or in the effective mass of $0.01\,m_0$ changes the $10^{14}\,\mathrm{m}^{-2}$ confinement boundary by approximately $1$--$2$ pixels at most, corresponding to $1.5$--$3\,\mathrm{nm}$. Although these parameter variations also change the absolute charge density, the horizontal and vertical confinement orientations remain unchanged. The designed confinement regime is therefore insensitive to small variations in these parameters.

\subsection*{RET preprocessing}

\begin{figure*}[tp]

\centering

\includegraphics[width=\linewidth,height=0.72\textheight,keepaspectratio]{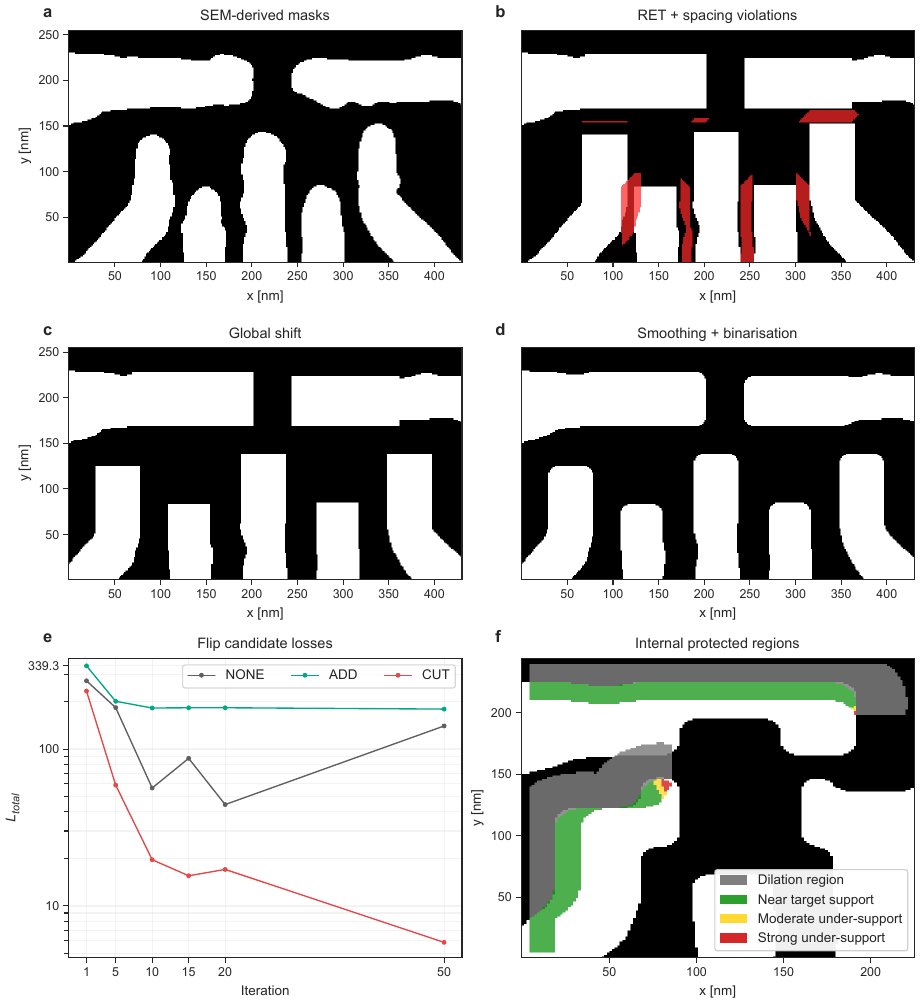}

\caption{\textbf{Electrode-geometry preprocessing and geometric decision terms.}
\textbf{a} SEM-derived electrode masks used as the starting geometry.
\textbf{b} Electrode masks after transformation of the selected dot-facing electrode edges into locally rectangular regions, with electrode-spacing violations highlighted in red.
\textbf{c} Electrode geometry after the global spacing correction.
\textbf{d} Final preprocessed geometry after Gaussian smoothing and binarisation.
\textbf{e} Candidate losses for the no-edit (NONE), addition (ADD) and cut (CUT) states evaluated at the scheduled flip checks for an example electrode edit.
\textbf{f} Construction of the internal fabrication loss $\mathcal{L}_{\mathrm{int}}$, showing the auxiliary dilation regions in grey and the electrode-mask values within the protected region. Green pixels lie above the threshold and contribute no loss, yellow pixels have intermediate electrode-mask values, and red pixels have the lowest electrode-mask values and therefore the largest contribution to $\mathcal{L}_{\mathrm{int}}$.}

\label{fig:5}

\end{figure*}

The rectangular edge transformation (RET) provides the geometry-processing and parameterisation framework used to prepare the SEM image of the gate electrode layout and subsequently modify selected gate boundaries during inverse design. Before optimisation, RET preprocesses the gate geometry extracted from the SEM image so that the modifiable gate electrode edges have a consistent local representation and the initial layout satisfies the imposed fabrication constraints. The individual electrodes are extracted from the SEM image as separate binary images, illustrated in Fig.\ref{fig:5}a, and calibrated using the scale bar of the device image in Ref.~\cite{Jirovec2021}, corresponding to an image scale of $1.486\,\mathrm{nm}$ per pixel.

The first preprocessing step transforms the selected dot-facing edges of the gate electrodes into locally rectangular regions. For each target QD, the surrounding gate electrodes are assigned left, right, top or bottom dot-facing edges. The entire modifiable gate electrode end is regularised such that the dot-facing boundary has a constant $x$- or $y$-coordinate, depending on its orientation, while the remainder of the electrode remains unchanged. This removes local irregularities inherited from the SEM-derived geometry and provides a well-defined rectangular boundary for subsequent additions and cuts, fabrication-constraint evaluation and geometry optimisation \cite{Luce2024dCG}. The fixed edge coordinate also ensures that modifications defined by $R_{\mathrm{sp}}$, $R_l$ and $R_d$ remain continuously attached to the original gate electrode as their parameters change, avoiding gaps or uneven overlap between the electrode and the parameterised modification. The transformation is illustrated in Fig. \ref{fig:5}b.

Following the gate electrode-edge transformation, the second preprocessing step enforces a global minimum gate-to-gate electrode separation of 10 pixels, corresponding to approximately 15 nm. Each electrode is dilated by half of the required separation and the dilated masks are checked for overlap. An overlap identifies gate electrode pairs that are closer than the specified design margin, and the corresponding gate electrodes are shifted apart until the overlap is removed. The correction is applied to the complete gate electrode layout, including regions outside the local inverse-design windows, because the starting masks are extracted from a fabricated device in which some gate pairs are closer than the desired design margin. This produces a starting geometry that satisfies the chosen minimum electrode separation across the full device and ensures that the final inverse-designed layout remains compatible with the imposed fabrication requirements. The corrected geometry is shown in Fig. \ref{fig:5}c.

The final preprocessing step smooths the gate electrode boundaries. Rectangular additions, cuts and locally regularised gate electrode ends contain ideal sharp corners that do not represent the rounded boundaries of fabricated electrodes. We therefore apply a Gaussian filter to the complete gate electrode mask, producing a continuous smoothed representation of the boundary, and subsequently threshold the result to recover a binary gate electrode mask. The Gaussian width $\sigma$ controls the spatial extent of the smoothing and therefore the degree of rounding applied to the gate electrode boundary. Applying the filter to the complete gate electrode ensures that the original boundary and the modified region are treated consistently. For the inverse-design calculations, we use $\sigma=4$ pixels, corresponding to approximately 6 nm, which rounds the gate electrode boundaries while limiting erosion or displacement of narrow gate electrode features during thresholding. The final smoothed and binarised geometry is shown in Fig.\ref{fig:5}d.

\subsection*{The Fourier Neural Operator}

Neural operators \cite{Azizzadenesheli2024,Kovachki2023,Berner2026} provide a principled extension of neural networks for learning mappings between function spaces. They approximate highly nonlinear operators by composing linear integral operators $\mathcal{K}$ with pointwise nonlinear activation functions $\sigma$, and possess a universal operator approximation property \cite{Kovachki2021_2}. More specifically, we define the neural operator
\begin{equation}
\mathcal{G}_{\phi}=\mathcal{Q}\circ\sigma(W_L+\mathcal{K}_L+b_L)\circ\cdots\circ\sigma(W_1+\mathcal{K}_1+b_1)\circ\mathcal{P},
\end{equation}
where $\mathcal{P}$ and $\mathcal{Q}$ are pointwise neural networks that lift a lower-dimensional function to a higher-dimensional representation and project it back. The model consists of $L$ stacked layers of the form $\sigma(W_l+\mathcal{K}_l+b_l)$, where $W_l$ are pointwise linear operators, $\mathcal{K}_l$ are integral kernel operators, $b_l$ are bias terms and $\sigma$ denotes a fixed activation function. The parameter set $\phi$ contains all trainable parameters in $\mathcal{P}$, $\mathcal{Q}$, $W_l$, $\mathcal{K}_l$ and $b_l$.

Several neural operator architectures have been introduced, including the Fourier Neural Operator (FNO) \cite{Li2021FNO,Duruisseaux2025FNO}. An FNO is a neural operator whose integral operator layers are defined in Fourier space as
\begin{equation}
\bigl(\mathcal{K}(\phi)v_t\bigr)(x)=\mathcal{F}^{-1}\!\left(R_\phi\cdot(\mathcal{F}v_t)\right)(x),
\end{equation}
where $R_\phi$ is the Fourier transform of a periodic function $\kappa$ parameterised by $\phi$. On a uniform mesh, the Fourier transform $\mathcal{F}$ can be computed efficiently using the fast Fourier transform (FFT). Prior knowledge of physical laws and governing differential equations can also be incorporated during training through additional loss terms, either supplementing or replacing reference data, as in physics-informed neural operators \cite{Li2024PINO,Lin2025mGNO,Ganeshram2025FCPINO}.

Within each FNO layer, latent functions are transformed into Fourier space, learned transformations are applied to a finite set of Fourier modes, and the resulting representation is transformed back to the original domain. Since Fourier modes have global support, these spectral operations exchange information across the full domain and provide a global receptive field suited to the non-local electrostatic response of gate-defined QD devices.

\subsection*{FNO dataset generation and training}

We generate the FNO training data using the physics-based electrostatic and Thomas--Fermi solver described above, while keeping material-specific parameters fixed. To construct the geometry dataset, we start from the gate electrode masks extracted from the SEM image. As a fixed preprocessing step, the selected dot-facing gate electrode edges are first transformed into rectangles as described in the RET subsection. This transformation is applied consistently before any random geometry variations are introduced.

After transforming the selected gate electrode edges into rectangles, we first vary the global gate electrode spacing from 2 to 16 pixels, corresponding to approximately 3--24 nm, generating layouts with different separations between neighbouring gate electrodes. Dataset geometries are then generated using sequential rectangular gate electrode modifications. For each QD region, we randomise the order of the four surrounding gate electrodes. Before modifying each gate electrode, we determine the admissible modification region from the current gate geometry, the QD position and the previously modified neighbouring gates. We then select either if the gate electrode should expand or contract. When both operations satisfy the geometric constraints, the operation is selected randomly. When only one is admissible, that operation is used. The rectangle position $R_{\mathrm{sp}}$, length $R_l$ and depth $R_d$ are subsequently sampled from the remaining admissible parameter space. Both $R_l$ and $|R_d|$ are constrained to be at least 7 pixels, corresponding to approximately 10.5 nm, preventing narrow expansions or contractions that would create gate electrodes that are difficult to fabricate. Their upper bounds depend on the available local geometry. After all gate electrode modifications are applied, we accept the geometry only if every electrode remains connected and the required gate-to-gate separation is satisfied. Otherwise, the sample is regenerated.

Finally, Gaussian smoothing is applied to the modified geometry. The smoothing width $\sigma$ is sampled from 0.5 to 10 pixels in increments of 0.5 pixels, exposing the FNO to different degrees of gate-edge rounding and to geometries representative of fabrication-compatible rounded boundaries.

Gate voltages are sampled independently of geometry using three complementary regimes to provide both dense coverage around the experimentally established double-QD configuration and broader coverage of the accessible gate voltage space. The narrowest regime samples each gate within a window of approximately 0.8~V on average (0.4--1.5~V depending on the gate electrode) around the experimentally reported voltages of Ref.~\cite{Jirovec2021}. An intermediate regime widens these windows to approximately 1.9~V on average (1.5--2.5~V), while the third regime samples every gate voltage uniformly over the full 0--4~V interval. The first two regimes increase the representation of configurations close to that used in Ref.\cite{Jirovec2021}. The full-range regime also contains many gate voltage combinations that would be unrealistic for a working device. They are included deliberately so that the FNO stays accurate over the whole gate voltage space, including the configurations the inverse-design optimisation passes through as it moves away from the starting point towards the target charge density.

The original SEM image spans $934\times625$ pixels. Simulating the full field is unnecessary for the double-QD training data because regions sufficiently far from the double-QD have negligible influence on its electrostatic potential. We therefore crop the physics-based simulation domain to $343\times399$ pixels, retaining the gate electrodes and surrounding region relevant to the double-QD confinement. A typical self-consistent charge density calculation on the cropped domain requires approximately 24~s, whereas the corresponding calculation on the original $934\times625$ domain requires approximately 153~s. 

The physics-based solver produces gate-potential and charge density fields on the $343\times399$ simulation grid. Before constructing the FNO inputs and target charge density maps, we remove five pixels from each boundary, resulting in a $333\times389$ FNO domain. Charge density values close to the simulation boundary can contain large edge artefacts that are unrelated to the double-QD confinement region. Excluding these pixels prevents the FNO from learning these non-physical boundary features and focuses the training objective on the region relevant to the device.

Using the sampled geometries and gate voltages, we generate 30,000 training configurations and 6,000 independent test configurations. Before training, each geometry and its corresponding gate voltages are combined into a single spatial input field. For each gate electrode $i$, $G_i(x,y)\in\{0,1\}$ denotes its binary spatial mask. Because the gate electrode masks do not overlap, the background mask is
\begin{equation}
B(x,y)=1-\sum_iG_i(x,y).
\end{equation}
The gate voltage is normalised as $V_i^{\mathrm{norm}}=V_i/(10\,\mathrm{V})$, so the 0--4 V operating range maps to values between 0 and 0.4. The single-channel FNO input is
\begin{equation}
a(x,y)=\sum_iG_i(x,y)V_i^{\mathrm{norm}}-B(x,y).
\end{equation}
Thus each gate electrode pixel carries the uniform normalised gate voltage of its gate electrode, while background pixels have the value $-1$. The model output is the predicted charge density on the same spatial grid.

Kossaifi et al.~\cite{Kossaifi2024Neural} provide a comprehensive open-source PyTorch library for learning neural operators, which serves as the basis of our implementation. We performed a hyperparameter search over all combinations of 64, 96 and 128 retained Fourier modes, 12, 20 and 28 hidden channels, and three and five Fourier layers. The selected model uses 64 retained Fourier modes, 20 hidden channels and three Fourier layers, which yielded the lowest error among the evaluated architectures. The model is trained for 200 epochs with the AdamW optimiser \cite{Loshchilov2019}.

\subsection*{Gradient propagation and optimisation}

During inverse design, the trained FNO parameters remain fixed and only the device-design variables are updated. Each selected dot-facing gate edge is described by variables in the design vector $\boldsymbol{\theta}=(R_{\mathrm{sp}},R_l,R_d,V)$. For a given $\boldsymbol{\theta}$, the parameterised gate geometry and voltages define the FNO input and corresponding predicted density $\rho_{\mathrm{FNO}}(\boldsymbol{\theta})$. The optimal design is
\begin{equation}
\boldsymbol{\theta}^{*}
=
\operatorname*{arg\,min}_{\boldsymbol{\theta}}
\mathcal{L}\!\left(
\rho_{\mathrm{FNO}}(\boldsymbol{\theta}),
\boldsymbol{\theta}
\right).
\end{equation}

During optimisation, $t$ denotes the optimisation iteration, $\boldsymbol{\theta}^{(t)}$ denotes the design parameters at iteration $t$ and $\rho^{(t)}$ denotes the corresponding FNO-predicted charge density. Because the FNO and continuous geometry construction provide a differentiable path to the design variables, gradients of the total loss can be propagated to the geometry and voltage parameters. A generic gradient-based update is
\begin{equation}
\boldsymbol{\theta}^{(t+1)}
=
\boldsymbol{\theta}^{(t)}
-
\eta
\nabla_{\boldsymbol{\theta}}
\mathcal{L}\!\left(
\rho^{(t)},
\boldsymbol{\theta}^{(t)}
\right),
\end{equation}
where $\eta$ denotes the learning rate. In the implementation, the continuous variables are updated with the Adam optimiser \cite{Kingma2015}, using separate learning rates for $R_{\mathrm{sp}}$, $R_l$, $R_d$ and $V$. The geometry parameters are represented by unconstrained optimisation variables and mapped differentiably into their allowed physical ranges. The gate voltages are optimised directly in millivolts and clamped to the allowed range. The learning rates follow a cosine-decay schedule over the optimisation.

The inverse design is initialised from the double-QD configuration in Ref.~\cite{Jirovec2021}. The trainable gate voltages begin at the reported values, while the geometry parameters are initialised close to the regions where the target confinement requires the gate edges to change. For each selected edge, the centre of the initial modification is aligned with the principal axis of the target ellipse. The tangential length $R_l$ is initialised slightly above its minimum allowed value, chosen as a hyperparameter for the results presented here. This places the initial modification in the region expected to have the strongest influence on the target charge density boundary. The depth $R_d$ is the parameter that primarily determines how far the gate electrode edge moves towards or away from the QD and therefore requires the largest adjustment during optimisation. Because its optimal value is not known from the target geometry alone, $|R_d|$ is initialised substantially above the minimum allowed depth and is subsequently refined by gradient optimisation.

The gate electrodes can undergo three discrete operations. Expansion extends the gate-electrode edge towards the QD, contraction retracts it away from the QD, and no edit leaves the gate-electrode geometry unchanged. These operations are referred to as addition, cut and no-edit states, respectively. Although continuous transitions between expansion and contraction are possible, they could result in final geometries containing modifications that are too small to fabricate reliably. We therefore introduce scheduled discrete operation evaluations, referred to as flip checks, during which the three states are compared and the preferred operation is selected.

Each selected edge retains its own $R_{\mathrm{sp}}$, $R_l$ and $R_d$ parameters. A gate electrode shared by two QDs can have separate local transformations acting on the gate edges facing each dot during simultaneous optimisation, while retaining a single shared trainable gate voltage. When these transformations contain overlapping additions and cuts, cuts are applied before additions. Overlapping addition and cut operations do not commute, so applying them in a fixed order ensures that a given set of transformation parameters always produces the same final gate electrode geometry.

At each optimisation iteration, the continuous geometry parameters are mapped into their allowed local ranges using differentiable bounded parameterisations that enforce the minimum tangential length and minimum non-zero depth. We use a straight-through estimator (STE) to preserve a binary gate electrode representation in the forward calculation while supplying a smooth derivative during backpropagation. The geometry pipeline uses the STE at two stages because a binary gate electrode representation is required before and after Gaussian smoothing.

First, each local rectangle is applied as a hard binary addition or cut in the forward pass, while the backward pass follows smooth sigmoid functions placed at the same rectangle edges. The hard forward operation makes the modified region unambiguously gate electrode or background before smoothing, while the sigmoid representation supplies derivatives with respect to $R_{\mathrm{sp}}$, $R_l$ and $R_d$. Passing a soft rectangle directly into the smoothing step would leave fractional values at the modification boundary and cause the original gate electrode and new modification to be smoothed differently.

After all local modifications acting on a gate electrode have been combined, the complete modified electrode is Gaussian-smoothed. This produces fractional values at the rounded gate boundary. A second STE binarises the smoothed gate electrode in the FNO forward pass while retaining the gradient path through smoothing. This ensures that every pixel assigned to a gate electrode receives the same gate voltage. Without this second binarisation, fractional boundary values would multiply the gate voltage and create an artificial gate voltage variation within one electrode. The resulting binary gate masks are combined with the normalised gate voltages and the $-1$ background value to form the FNO input $a(x,y)$.

All selected dot-facing edits are initially placed in the no-edit state. This allows the continuous gate voltages to undergo an initial Adam optimiser update before the first flip check, so that the electrostatic confinement can first respond through gate voltage changes before a discrete geometry modification is selected. The geometry parameters of the no-edit state are retained for subsequent expansion or contraction of the gate electrode.

The discrete addition, cut and no-edit operations are re-evaluated at scheduled flip checks. Between flip checks, the selected operation for each edit remains fixed while its continuous geometry parameters and the participating gate voltages are updated by the Adam optimiser. At each flip check, the addition, cut and no-edit states are evaluated for each gate edit from the same state, and the state producing the lowest local loss is identified. When more than one gate electrode surrounding the same QD is proposed to change operation, the proposed changes are subsequently evaluated together. Each proposed gate electrode can either retain its current state or adopt its newly selected state, giving at most $2^4$ combinations for the four gate electrodes surrounding one QD. The combination producing the lowest loss is retained, preventing individually favourable changes from being committed together when their combined effect is unfavourable. For a gate electrode shared by two QDs, the charge density losses of both QDs are included when evaluating its candidate operations. Figure \ref{fig:5}e shows an example of total loss for each operation evaluated during the flip checks for a single gate electrode.

The no-edit state corresponds to effective values of \(R_l=R_d=0\). When the no-edit state is selected at a flip check because it produces the lowest loss, the geometry parameters \(R_{\mathrm{sp}}\), \(R_l\) and \(R_d\) receive no gradients from the main optimisation and therefore remain unchanged until the next flip check. Consequently, the addition and cut alternatives would be re-evaluated using the same geometry parameters as at the previous flip check, despite the gate voltages having evolved. This limits the ability of these alternatives to adapt their geometry towards the target confinement and may prevent the selection of modifications that would otherwise improve the result. To overcome this limitation, we introduce a parallel optimisation branch, referred to as \textit{shadow geometry}, whenever the no-edit state is selected. At each flip check, the addition and cut alternatives are evaluated, and the one producing the lower loss is retained as the shadow operation. This alternative is then optimised in parallel with the no-edit state using the same gate voltages, which are detached from the shadow gradient calculation. The resulting gradients update only the corresponding \(R_{\mathrm{sp}}\), \(R_l\) and \(R_d\) parameters, allowing the potential modification to evolve without affecting the currently selected gate geometry or the gate voltage optimisation. At subsequent flip checks, the addition and cut alternatives are therefore evaluated using updated geometry parameters that have adapted to the evolving voltage configuration.

The flip checks are restricted to the initial stages of optimisation, during which the gate geometry and voltages undergo their largest adjustments and the preferred discrete operations are established. As the optimisation progresses, the selected operations are expected to stabilise, while subsequent iterations predominantly refine the continuous geometry parameters and gate voltages. After the final scheduled flip check, the addition, cut or no-edit state of each edit is therefore fixed for the remainder of the optimisation. This also reduces the computational cost, as the alternative operation states and shadow geometries no longer need to be evaluated. The remaining iterations refine the continuous parameters of the selected geometry and the trainable gate voltages without further changes to the discrete operation states. After the prescribed number of optimisation iterations, the state with the lowest total loss recorded during the run is retained as the final design. Numerical parameter bounds, initial learning rates, cosine-decay settings, loss weights, STE parameters and the operation-check schedule are listed in Supplementary Table~S2.

\subsection*{Loss implementation}

The total objective is defined in the main text in terms of the density losses $\mathcal{L}_{\mathrm{in}}$ and $\mathcal{L}_{\mathrm{out}}$ and the fabrication losses $\mathcal{L}_{\mathrm{ext}}$ and $\mathcal{L}_{\mathrm{int}}$. Here we describe how these four terms are evaluated and how their gradients enter the optimisation.

Because the predicted charge density varies continuously, we define the QD region using the hard charge density threshold $\rho_{\mathrm{th}}=10^{14}\,\mathrm{m}^{-2}$. In the forward pass, the binary field is
\begin{equation}
o(x,y)=\begin{cases}
1,&\rho_{\mathrm{FNO}}(x,y)\geq\rho_{\mathrm{th}},\\
0,&\text{otherwise}.
\end{cases}
\end{equation}
For the prescribed target ellipse $\Omega_{\mathrm{in}}$ and the surrounding rectangle with the target ellipse removed, $\Omega_{\mathrm{out}}$, the mean field values are
\begin{equation}
\begin{aligned}
m_{\mathrm{in}}&=\frac{1}{|\Omega_{\mathrm{in}}|}\sum_{(x,y)\in\Omega_{\mathrm{in}}}o(x,y),\\
m_{\mathrm{out}}&=\frac{1}{|\Omega_{\mathrm{out}}|}\sum_{(x,y)\in\Omega_{\mathrm{out}}}o(x,y).
\end{aligned}
\end{equation}
The corresponding density losses are
\begin{equation}
\begin{aligned}
\mathcal{L}_{\mathrm{in}}&=\left[\max\!\left(0,\alpha_{\mathrm{in}}-m_{\mathrm{in}}\right)\right]^2,\\
\mathcal{L}_{\mathrm{out}}&=\left[\max\!\left(0,m_{\mathrm{out}}-\beta_{\mathrm{out}}\right)\right]^2.
\end{aligned}
\end{equation}
The thresholds $\alpha_{\mathrm{in}}$ and $\beta_{\mathrm{out}}$ define the minimum acceptable fraction of pixels above $\rho_{\mathrm{th}}$ in the inner region and the maximum acceptable fraction in the outer region. Once the corresponding fraction satisfies its threshold, that loss becomes zero. Filling $\Omega_{\mathrm{in}}$ while emptying $\Omega_{\mathrm{out}}$ drives the charge density boundary towards the prescribed ellipse and controls both its orientation and lateral extent.

The hard charge density threshold is non-differentiable. During backpropagation, a separate STE is used for the charge density field. The forward loss is evaluated from the hard threshold above, while the backward pass uses the derivative of a smooth threshold centred at $\rho_{\mathrm{th}}$. Gradients from $\mathcal{L}_{\mathrm{in}}$ and $\mathcal{L}_{\mathrm{out}}$ propagate through the fixed FNO to the gate-weighted input. From the input, the gradient follows a direct path to the trainable gate voltages and a geometry path through the second STE, Gaussian smoothing and the first STE at the rectangle edges rectangle edges to $R_{\mathrm{sp}}$, $R_l$ and $R_d$.

The external fabrication term enforces the minimum separation between different gate electrodes. For a selected gate $i$, the neighbouring gate electrodes are dilated by the required fabrication spacing to define the external protected mask $M_{\mathrm{ext}}^{(i)}(\mathbf{x})$. Let $G_i^{(t)}(\mathbf{x})$ denote the current differentiable map of the selected gate electrode at iteration $t$. The external loss is
\begin{equation}
\mathcal{L}_{\mathrm{ext}}=\sum_i\sum_{\mathbf{x}}M_{\mathrm{ext}}^{(i)}(\mathbf{x})G_i^{(t)}(\mathbf{x}).
\end{equation}
The external loss is evaluated on the differentiable smoothed gate electrode before the final binarisation used for the FNO input. An addition approaching the protected region around a neighbouring gate electrode therefore receives a continuous geometry gradient before the minimum separation is violated. Gaussian smoothing produces small non-zero values outside the binary gate electrode edge. These values would otherwise create a residual external loss even when the gate electrode remains outside the protected region. When evaluating $\mathcal{L}_{\mathrm{ext}}$, pixels with values at or below 0.05 in the smoothed gate-electrode masks are treated as zero. This removes the contribution caused only by edge smoothing while retaining the increasing values of the gate-electrode mask produced when the rectangular modification of the selected gate electrode approaches the protected region of a neighbouring gate electrode.

The internal fabrication term enforces the minimum remaining gate electrode width. A strict internal protected band is defined within each selected gate electrode, specifying a region of the gate geometry that must be preserved as a cut moves inward from the QD-facing edge. Let $M_{\mathrm{int}}^{(i)}(\mathbf{x})$ denote the internal protected mask and let $G_{i,\mathrm{int}}^{(t)}(\mathbf{x})$ denote the reconstructed differentiable gate-electrode mask used specifically for this constraint. The internal loss is
\begin{equation}
\mathcal{L}_{\mathrm{int}}=\sum_i\frac{1}{\sum_{\mathbf{x}}M_{\mathrm{int}}^{(i)}(\mathbf{x})}\sum_{\mathbf{x}}M_{\mathrm{int}}^{(i)}(\mathbf{x})\left[1-G_{i,\mathrm{int}}^{(t)}(\mathbf{x})\right].
\end{equation}
The loss is thus the sum over the average errors of the edits of $G_{i,\mathrm{int}}^{(t)}$ from within the protected band. A separate gate-electrode mask is required because ordinary Gaussian smoothing reduces the continuous value at every gate electrode edge even when the binary geometry is valid. Evaluating $\mathcal{L}_{\mathrm{int}}$ directly on the normally smoothed gate electrode would therefore produce a baseline penalty near the protected edge, even when the gate geometry remains valid.

For each active edit, the gate electrode produced by the normal geometry pipeline is first Gaussian-smoothed and thresholded to recover a clean binary geometry. The binary gate electrode is then scanned line by line. Rows are used for left and right edits and columns are used for top and bottom edits. On each line, the continuous gate electrode segment is identified and the internal protected band is marked on the side opposite the active modification. An auxiliary dilation region is generated from the gate electrode on this opposite side. It extends away from the active edge and across the transverse extent of the protected band. The auxiliary region is combined with the thresholded gate electrode and the reconstructed geometry is Gaussian-smoothed a second time. This produces the continuous field $G_{i,\mathrm{int}}^{(t)}$, which is used only for $\mathcal{L}_{\mathrm{int}}$.

When the cut remains far from the protected band, the auxiliary dilation keeps the reconstructed values inside the band close to one despite rounding introduced by the second smoothing step. As the cut approaches the protected region, the values of the pixels nearest the cut decrease continuously and $\mathcal{L}_{\mathrm{int}}$ increases, providing a gradient that redirects the optimisation before the minimum remaining gate electrode width is violated. The second Gaussian smoothing can leave otherwise valid protected pixels with reconstructed values slightly below one. When evaluating $\mathcal{L}_{\mathrm{int}}$, reconstructed values at or above 0.95 are therefore treated as one. This removes the residual loss caused only by smoothing while retaining the decrease in pixel values produced when a cut genuinely approaches the protected region. The construction of the internal protected region, auxiliary dilation and resulting differentiable gate geometry is illustrated in Fig. \ref{fig:5}f.

For each iteration, the current continuous parameters and discrete operation states define the gate geometry. The geometry is converted to the FNO input and the FNO predicts $\rho_{\mathrm{FNO}}$. The charge density and fabrication losses are then evaluated. Gradients from $\mathcal{L}_{\mathrm{in}}$ and $\mathcal{L}_{\mathrm{out}}$ propagate through the FNO, while $\mathcal{L}_{\mathrm{ext}}$ and $\mathcal{L}_{\mathrm{int}}$ depend directly on the differentiable gate electrode geometry and do not pass through the FNO. The charge density and fabrication gradient contributions are combined before the Adam optimiser updates the continuous geometry parameters and participating gate voltages in the same optimisation step.

\bibliographystyle{\arxivstylepath revtex_arxiv}

\bibliography{\arxivbibfile}

\section*{Code availability}
The code will be made publicly available on GitHub upon publication.

\section*{Acknowledgements}
We thank D. Jirovec for valuable discussions regarding the device fabrication and measurements, and operating conditions. We also thank Brucek Khailany for discussions on semiconducting materials and their simulation and inverse design, as well as J. T. Luna, A.R. Akhmerov, F. Fedele, and P. Vaidhyanathan for helpful discussions regarding the inverse-design algorithm and its implementation.
This work was supported by the European Research Council (grant agreement 948932), the Royal Society (URF-R1-191150), and the United States Army Research Office under Award No. W911NF-24-2-0043. Views and opinions expressed are however those of the authors only.

\section*{Competing interests}
N.A. is a founder of QuantrolOx, which develops machine learning-based software for quantum control. The other authors declare no competing interests.

\end{document}